\documentclass{article}
\usepackage{spconf,amsmath,graphicx}
\usepackage{multirow,amsfonts,cite,array,url}
\usepackage{color}
\usepackage{cuted}
\usepackage{stfloats}
\usepackage{enumitem}
\usepackage{subfigure}

\title{Transferring Structure Knowledge: A New Task to Fake news Detection Towards Cold-Start Propagation}

\name{
    Lingwei Wei$^{1,2}$ \qquad 
    Dou Hu$^{1,2}$ \qquad
    Wei Zhou$^{1*}$ \qquad
    Songlin Hu$^{1,2}$ 
    \thanks{* Corresponding author}
    }
\address{
    $^1$ Institute of Information Engineering, Chinese Academy of Sciences  \\
    $^2$ School of Cyber Security, University of Chinese Academy of Sciences \\ 
    \{weilingwei, hudou, zhouwei, husonglin\}@iie.ac.cn
}

\begin{document}
%
\maketitle
\begin{abstract}
Many fake news detection studies have achieved promising performance by extracting effective semantic and structure features from both content and propagation trees. However, it is challenging to apply them to practical situations, especially when using the trained propagation-based models to detect news with no propagation data. Towards this scenario, we study a new task named cold-start fake news detection, which aims to detect content-only samples with missing propagation. To achieve the task, we design a simple but effective Structure Adversarial Net (SAN) framework to learn transferable features from available propagation to boost the detection of content-only samples. SAN introduces a structure discriminator to estimate dissimilarities among learned features with and without propagation, and further learns structure-invariant features to enhance the generalization of existing propagation-based methods for content-only samples. We conduct qualitative and quantitative experiments on three datasets. Results show the challenge of the new task and the effectiveness of our SAN framework.
\end{abstract}

\section{Introduction}

Nowadays, mainstream social platforms (e.g., Twitter) have facilitated the dissemination of information in a faster and cheaper way.
Nevertheless, the ease has also caused the wide spread of {fake news}, which has brought detrimental effects on individuals and society \cite{loomba2021measuring}. 
Triggered by the negative impact of fake news spreading, it is critical to develop automatic methods for fake news detection.

Generally, users on social media share opinions, conjectures and evidence  for checking fake news. Through their various interactive behaviors, a propagation tree describing the law of information transmission is formed and plays a significant role in fake news detection. 
Previous works  \cite{Vosoughi1146,DBLP:journals/chb/JangGLXHKT18} have empirically shown that compared with the truth, false news has deeper propagation structures, and reaches a wider audience. To leverage the difference, many efforts \cite{DBLP:conf/acl/WongGM18,DBLP:conf/acl/KumarC19,DBLP:conf/coling/MaG20,dou2021rumor,DBLP:conf/aaai/BianXXZHRH20,DBLP:conf/acl/WeiHZYH20,DBLP:conf/emnlp/LinMCYCC21,wei2022modeling,DBLP:conf/coling/Wei00H22,DBLP:conf/coling/Wei0L0H22} have been denoted to jointly explore effective high-level semantic and structural properties from content and the corresponding propagation trees via different neural networks. Compared with the model learned only on the content \cite{DBLP:conf/ijcai/MaGMKJWC16,DBLP:conf/cikm/RuchanskySL17,karimi-tang-2019-learning}, these propagation-based models trained on samples with both content and propagation, provides a more comprehensive view of fake news and have shown superior detection performance.

However, a practical barrier is that, in most cases, the acquisition of propagation data is not available at any time, and usually requires a great quantity of manpower and computation resources. When lacking propagation structure, the above propagation-based detection systems would obtain suboptimal performance. These models are trained on content and propagation tree to jointly learn semantic and structural features, leading to a specific feature space for detection. Obviously, for samples that lack propagation information, i.e., \textit{cold-start propagation}, they fail to perform well due to the dissimilarities among features with and without propagation. 

Based on the above scenario, we develops a new task to fake news detection towards cold-start propagation, named \textbf{cold-start fake news detection}.
It aims to train the model from samples with available propagation and content, and then predict content-only samples without any propagation data. 
Different from the existing fake news detection tasks, the new task focuses on the model's generalization capability of absence of propagation trees.
Studying cold-start propagation scenario can promote the extensive applications of propagation-based detection methods in practical detection.

Under cold-start fake news detection task, directly applying existing propagation based models to capture structure-specific features from propagation trees would hurt the detection of the cold-start news that has no propagation trees. 
Therefore, an intuitive solution to the new task is to remove the nontransferable structure-specific features and preserve the shared characteristics across different data types.

To achieve this, we design a simple but effective \textbf{Structure Adversarial Net (SAN)} framework to boost the detection performance of propagation-based models for cold-start news. SAN is a unified framework and can be employed to any propagation-based model.
Specifically, inspired by \cite{NIPS2014_5ca3e9b1}, SAN incorporates a structure discriminator to predict whether the high-level representation of the target news includes structure properties.
During the training phase, the original feature extractor cooperates with the fake news detector to carry out identifying fake news. Simultaneously, the feature extractor tries to fool the structure discriminator to close the gap distributions of the presence and absence of propagation trees.

We conduct experiments on three public fake news benchmarks and build two different cold-start propagation settings (i.e., general and event-aware) to simulate the real-world detection scenario.
Experimental results show different degrees of the degradation of existing propagation-based detection methods, demonstrating the challenge  of fake news detection towards cold-start propagation. 
We also evaluate our proposed SAN framework on several baseline models. The results prove that SAN consistently enhances the detection performance of these methods under the cold-start propagation condition.
We further discuss several potential directions to promote the development of cold-start fake news detection.

The main contributions can be summarized as follows: 
1) We develop a new task, fake news detection towards cold-start propagation, which focuses on the generalization of absence of the whole propagation. The goal of the task is to identify the cold-start news by exploiting previously available propagation and contents. It allows propagation-based methods to be applied to more practical detection scenarios. 
2) We propose a simple but effective Structure Adversarial Net (SAN) framework to transfer structure knowledge for content-only samples. It can be applied to any propagation-based models to promote the generalization for detecting cold-start news. 
3) We explore different settings for cold-start fake news detection on three datasets. Experiments demonstrate the challenge of new task and the effectiveness of the proposed framework.

\begin{figure}[t]
\centering
\includegraphics[width=\linewidth]{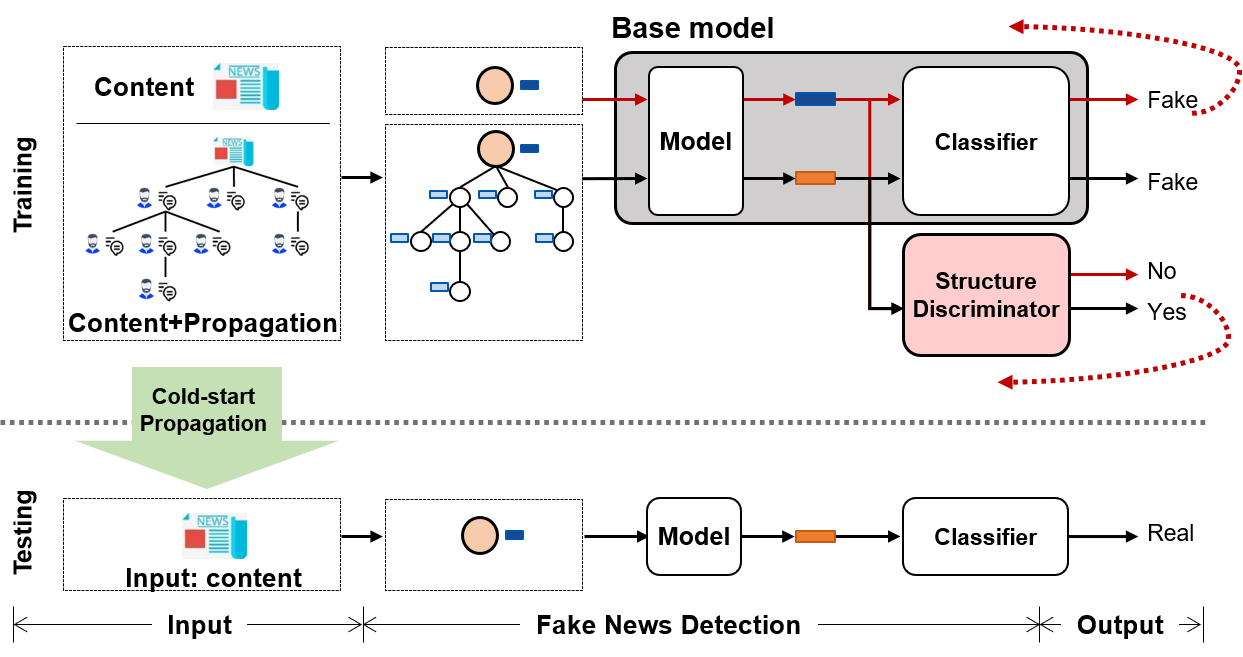}
\caption{The architecture of SAN framework for cold-start fake news detection. A structure discriminator is introduced to predict the auxiliary structure label based on the latent representation. SAN can transfer structure knowledge learned from existing propagation trees to content-only samples.}
\label{fig:model}
\end{figure}

\section{A New Task: Cold-start Fake news Detection}

We describe how to revise the traditional fake news detection to achieve the new task towards cold-start propagation.
Specifically, as shown in the Input in Fig.~\ref{fig:model}, we directly remove the whole propagation trees of samples in the testing set as the cold-start news to simulate the cold-start propagation setups. 
Formally, define $D^{\text{train}} = \{ {(x^{\text{train}}_i,G^{\text{train}}_i)}, i \in [1, N_{\text{train}}] \}$ and $D^{\text{test}} = \{ {(x^{\text{test}}_i)}, i \in [1, N_{\text{test}}] \}$ as the training and testing sets, respectively, where  $x$ refers to the source news and $G$ indicates propagation trees. 
During the training stage, the model is trained on the complete samples $(x^{\text{train}}_i,G^{\text{train}}_i)$ with both content and propagation to learn transferable semantic and structural patterns for detecting fake news, i.e., 
$$f: (x^{\text{train}},G^{\text{train}}) \rightarrow y.$$
During the testing stage, the trained model is used to predict the cold-start news $x^{\text{test}}_i$ that lacks propagation data, i.e., $$\hat{y} = f(x^{\text{test}}).$$

\section{Approach}
To boost the detection for content-only samples, we develop a simple but effective Structure Adversarial Net (SAN) framework to learn latent structural features from previous propagation trees. 
The overall architecture is shown in Fig.~\ref{fig:model}.

\subsection{Vanilla Propagation-based Approach}

Given the input sample including content of the source news $x$ and propagation trees $G$, existing models apply various neural networks to extract high-level textual and structural features.
The latent representation $\textbf{h}$ is computed by,
\begin{equation}
\textbf{h} = f_{\text{enc}}(x, G; \Theta),
\end{equation}
where $f_{\text{enc}}$ can be the encoder in \cite{DBLP:conf/aaai/BianXXZHRH20,DBLP:conf/acl/WeiHZYH20,DBLP:conf/coling/Wei00H22} to learn semantic and structural features, and $\Theta$ refers to the corresponding trainable parameters. 
Then, a classifier consisting of a full connection layer and a softmax function, is applied to predict the label probabilities of all classes, i.e.,
\begin{equation}
\hat {\textbf{y}} = f_{\text{cls}}(\textbf{h}; \theta_f),
\end{equation}
where $\theta_f$ is the classifier's learnable parameters.

\subsection{Structure Adversarial Net Framework}
As previous propagation-based detection methods fail to generalize well for content-only samples, we design a Structure Adversarial Net (SAN) framework to learn a transferable feature representation between content and propagation.

Based on the architecture of existing detection methods, SAN incorporates a structure discriminator to predict whether the high-level representation of the target news includes structure properties. Given the hidden representation, we leverage a full connection layer and a softmax function to predict label probabilities $\hat{\textbf{y}}_d$  of the representation computing from  the propagation structure, i.e., 
\begin{equation}
\hat {\textbf{y}}_d = f_{\text{d}}(\textbf{h}; \theta_d),
\end{equation}
where $\theta_d$ is learnable parameters of the classifier. ${\textbf{y}}_d = 1$ refers to the representation is learned from original samples with content and propagation; ${\textbf{y}}_d = 0$ refers to the representation is solely learned from content-only samples.

During the training, the original feature extractor cooperates with the fake news detector to carry out the major task of identifying fake news. The classification loss $L_{\text{CLS}}(\Theta, \theta_{f})$ is defined as, 
\begin{equation}
\mathcal{L}_{\text{CLS}}(\Theta, \theta_{f}) = - \textbf{y} \log (\hat{\textbf{y}}) - (1-\textbf{y}) \log (1 - \hat{\textbf{y}}),
\end{equation}
where $\textbf{y}$ is the ground-truth and $\hat{\textbf{y}}$ is prediction distribution.
Simultaneously, the feature extractor tries to fool the structure discriminator to close the gap across distributions from contents and propagation trees. 
The loss of the discriminator captures the dissimilarities of feature representations from different data types. It is defined as, 
\begin{equation}
\mathcal{L}_d (\Theta, \theta_{d}) = - \textbf{y}_d log(\hat{\textbf{y}}_d) + (1-\textbf{y}_d)log(1-\hat{\textbf{y}}_d),
\end{equation}
where $\textbf{y}_d$ and $\hat{\textbf{y}}_d$ are the ground-truth and prediction labels that describe whether the high-level representation of the target news includes structure properties, respectively. The larger the loss, the lower the dissimilarities. 
Cooperating with the fake news classifier $f_{\text{cls}}$ to minimize the cross-entropy loss, the final objective of optimization can be defined as,
\begin{equation}
\mathcal{L}_{\text{SAN}} = \mathcal{L}_{\text{CLS}}(\Theta, \theta_{f}) - \mathcal{L}_{\text{d}}(\Theta, \theta_{d}).
\end{equation}

The gradient reversal layer \cite{DBLP:conf/icml/GaninL15} is added between encoder and the structure discriminator to achieve an adversarial effect. 
Thus, the optimization of the model parameters are summarized as follows: 
\begin{equation}
\begin{split}
\Theta &\leftarrow \Theta - \eta (\frac{\partial \mathcal{L}_{\text{CLS}}} {\partial \Theta} - \frac{\partial \mathcal{L}_{\text{d}}} {\partial \Theta} ),  \\ 
\theta_f &\leftarrow \theta_f - \eta \frac{\partial \mathcal{L}_{\text{CLS}} }  {\partial \theta_f},   \theta_d \leftarrow \theta_d - \eta \frac{\partial \mathcal{L}_{d} }  {\partial \theta_d}, 
\end{split}
\end{equation}
where $\eta$ is the learning rate.
In the implementation, samples in the training set are processed into two copies. One is the full samples with both content and propagation, denoted as $D^{\text{train}}$, and the other only contains cold-start samples lacking the whole propagation, denoted as $\tilde{D}^{\text{train}}$.
We adopt the objectives of SAN for both of them, i.e., 
\begin{equation}
\mathcal{L} = \mathcal{L}_{\text{SAN}}^{D^{\text{train}}} + \lambda  \mathcal{L}_{\text{SAN}}^{\tilde{D}^{\text{train}}}, 
\end{equation}
where $\lambda$ is a trade-off hyper-parameter to control weights of considering cold-start samples during the training.

\section{Experiments}

\subsection{Experimental Setups}

\textbf{Datasets.} 
We experiments on three real-world public datasets.
{PolitiFact} and {GossipCop} are released by \cite{DBLP:journals/bigdata/ShuMWLL20}. Samples are collected from two fact-checking websites \textit{PolitiFact}\footnote{\url{https://www.politifact.com/}} and \textit{GossipCop}\footnote{\url{https://www.gossipcop.com/}}.
{PolitiFact} provides 157 fake news and 157 true news; {GossipCop} provides 2,732 fake news and 2,732 true news.
{PHEME-5} \cite{DBLP:journals/corr/ZubiagaHLPT15} contains tweets related to five different events. It contains 581 true news and 230 fake news.

\noindent \textbf{Task Setups.} 
We build two different cold-start propagation settings.
\textit{General cold start fake news detection} aims to detect content-only fake news without considering specific events. We choose PolitiFact and GossipCop, and follow the same procedure as \cite{DBLP:conf/kdd/ShuCW0L19,DBLP:conf/coling/Wei00H22} to split each dataset, {i.e.,} randomly choose 75\% of the data as the training set and keep the rest as the test set. 
We further remove the propagation trees and only retain the source news for each sample in the test set to achieve the cold-start propagation.
\textit{Event-aware cold start fake news detection} focus on detecting cold-start fake news for the new event. We evaluate on PHEME-5, which contains news from five specific events.
We use one event's samples are used for testing, and all the rest are used for training. Similarly, we further remove the whole propagation tree  for each sample in the test set.

\noindent \textbf{Baselines.}
\textbf{mGRU} \cite{DBLP:conf/ijcai/MaGMKJWC16} and \textbf{CSI} \cite{DBLP:conf/cikm/RuchanskySL17} are RNN-based models to capture sequential patterns from retweet sequences.
\textbf{GCNFN} \cite{DBLP:journals/corr/abs-1902-06673} models the propagation structure as a graph and uses graph convolutional networks (GCN) to encode the propagation. We implemented the model by removing profile information for a fair comparison. 
\textbf{GAT} \cite{DBLP:conf/iclr/VelickovicCCRLB18} applies graph attention networks to encode the propagation.
\textbf{BiGCN} \cite{DBLP:conf/aaai/BianXXZHRH20}  employs two GCNs to model the propagation graph and dispersion graph. 
\textbf{UPSR} \cite{DBLP:conf/coling/Wei00H22} is a state-of-the-art model that reconstructs latent propagation structure to explore more accurate and diverse structural properties. 
Besides, we also report a baseline that only using the content of source news for detection, denoted as \textbf{Content}, to evaluate the role of propagation trees on fake news detection tasks. We extracted textual features by word2vec embeddings and then fed them into the MLP for classification.

\noindent \textbf{Implementation Details.}
For PolitiFact and GossipCop, we use 300-dimensional word2vec vectors \cite{DBLP:journals/corr/abs-1301-3781} provided by \cite{DBLP:conf/sigir/DouSXYS21} as the input features of text contents.
For PHEME-5, we extract text embedding of each sentence by skip-gram with negative sampling \cite{DBLP:conf/nips/MikolovSCCD13}, and the dimension of input vectors is also set to 200. 
The dimension of hidden vectors is set to 64.
$\lambda$ is searched from $\{0.1, 1, 1.5, 2, 5, 10\}$.
The learning rate is set to 0.001, 0.0005, and 0.005 for {\it PolitiFact}, {\it GossipCop}, and PHEME-5.
We run each model with five random seeds and report the average results of the test set.

\begin{figure}[t]
\centering
\includegraphics[width=\linewidth]{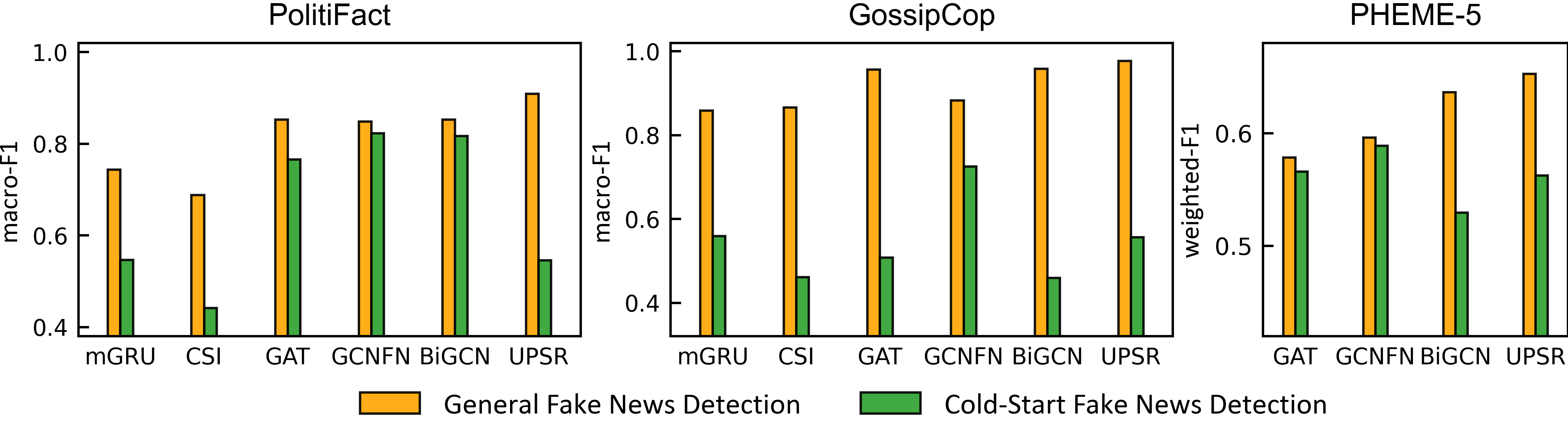}
\caption{Results on different fake news detection tasks. 
}
\label{fig:phe}
\end{figure}
\subsection{Task Analysis}
We first quantitatively evaluate existing propagation-based methods for two types of cold-start propagation.
Results are shown in Fig.~\ref{fig:phe}. 
For PolitiFact and GossipCop, we consider a mixture of social events; and for PHEME-5, we consider the event-specific setting.
From results, when direct applying propagation-based methods to the new task, the results of all comparison models decrease to a varying degree in terms of metrics. 
The inferior results show that these models do not generalize well to cold-start propagation.
For the same dataset, the more complex the models, the greater the performance degradation. It may be because the complex model is prone to fuse excessive nontransferable structure features from the propagation tree. Once the propagation structure is missing, the model cannot perform well.

\begin{figure}[t]
\centering
\includegraphics[width=\linewidth]{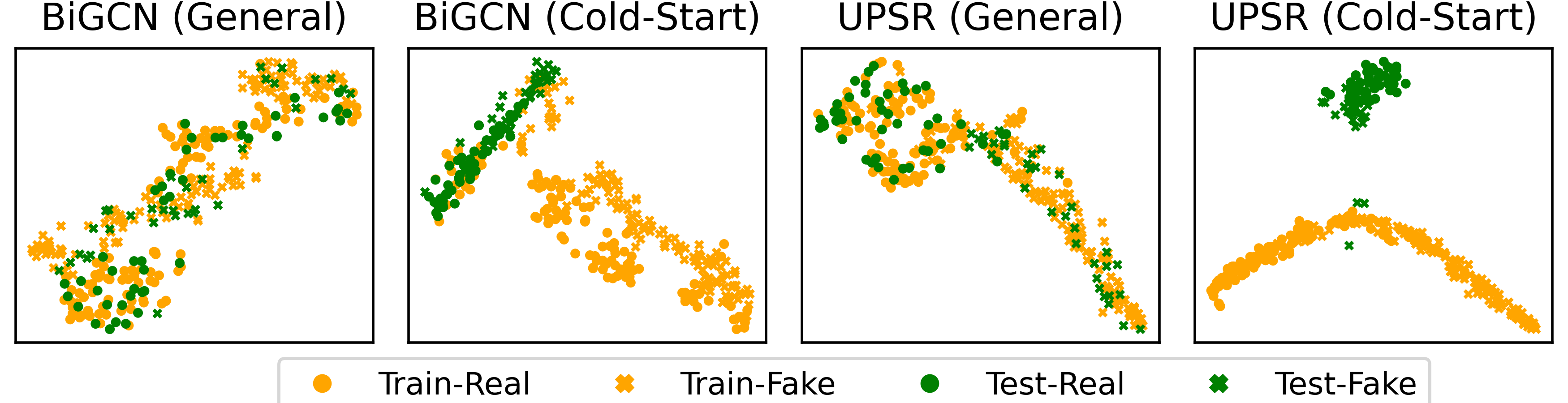}
\caption{Visualization of representations on PolitiFact.
}   
\label{fig:vis}
\end{figure}

Figure~\ref{fig:vis} qualitatively visualize the sample's representations on the training and testing set of PolitiFact with t-SNE \cite{van2008visualizing}.
We observe the inconsistent representation distribution of training samples and test samples. 
For the new task, it is critical to learn transferable patterns between content and propagation for the new task so that the model can adapt to detect content-only samples without propagation structure.

\begin{table}[t]
\centering
\resizebox{\linewidth}{!}{
\begin{tabular}{l|c|c|c|c|c|c|c|c}
\hline 
\multicolumn{1}{c|}{\multirow{3}{*}{Model}}  &
\multicolumn{4}{c|}{\multirow{1}{*}{PolitiFact}} &
\multicolumn{4}{c}{\multirow{1}{*}{GossipCop}}  \\ \cline{2-9}
& \multirow{2}{*}{Acc} 
& \multirow{2}{*}{ma-F1} 
& \multicolumn{2}{c|}{F1}
& \multirow{2}{*}{Acc}
& \multirow{2}{*}{ma-F1} 
& \multicolumn{2}{c}{F1}   \\  \cline{4-5} \cline{8-9}
& 
& &{\it fake} & {\it real} &
& & {\it fake} & {\it real}  \\ 
\hline 
\multicolumn{1}{l|}{Content}  
& 0.684 & 0.668 & 0.641 & 0.714 
& 0.755 & 0.749 & 0.756 & 0.753 \\    \hline
mGRU 
& 0.592 & 0.546 & 0.654 & 0.448 & 0.582 & 0.559 & 0.488 & 0.643
\\ 
\textbf{mGRU+SAN}   
& {\bf 0.699} & {\bf 0.660} & 0.629 & {\bf 0.719} & {\bf 0.710} & {\bf 0.702} & {\bf 0.721} & {\bf 0.696} \\ 
\textbf{Improve} (\%) & 
{\bf +10.7} & {\bf +11.4} & -2.5 & {\bf +27.1} & {\bf +12.8} & {\bf +14.3} & {\bf +23.3} & {\bf +5.3} \\
\hline
CSI             
&
0.562 & 0.442 & 0.337 & 0.562 & 0.552 & 0.461 & 0.256 & 0.677 \\ 
\textbf{CSI+SAN} 
& 
{\bf 0.610} & {\bf 0.558} & {\bf 0.563} & {0.565} & {\bf 0.705} & {\bf 0.697} & {\bf 0.691} & {\bf 0.714} \\ 
\textbf{Improve} (\%) & 
{\bf +4.8} & {\bf +11.6} & {\bf +22.6} & {+0.3} & {\bf +15.3} & {\bf +23.6} & {\bf +43.6} & {\bf +3.7} \\ \hline
GAT 
&
0.777 & 0.766 & 0.781 & 0.760 & 0.574 & 0.508 & 0.349 & 0.680 \\
\textbf{GAT+SAN} 
&
{\bf 0.851} & {\bf 0.844} & {\bf 0.848} & {\bf 0.848} & {\bf 0.751} & {\bf 0.745} & {\bf 0.750} & {\bf 0.752}     \\ 
\textbf{Improve} (\%) &
{\bf +7.3} & {\bf +7.7} & {\bf +6.7} & {\bf +8.8} & {\bf +17.8} & {\bf +23.7} & {\bf +40.1} & {\bf +7.1} \\ \hline  
GCNFN
& 0.830 & 0.823 & 0.829 & 0.823 & 0.731 & 0.725 & 0.727 & 0.735\\ 
\textbf{GCNFN+SAN} 
& 
{\bf 0.843} & {\bf 0.834} & {\bf 0.843} & {\bf 0.836} & {\bf 0.755} & {\bf 0.749} & {\bf 0.756} & {\bf 0.754} \\ 
\textbf{Improve} (\%) &  
{\bf +1.3} & {\bf +1.1} & {\bf +1.4} & {\bf +1.3} & {\bf +2.4} & {\bf +2.4} & {\bf +2.9} & {\bf +1.9} \\ \hline
BiGCN 
& 
0.825 & 0.817 & 0.834 & 0.808 & 0.559 & 0.460 & 0.243 & 0.688
\\ 
\textbf{BiGCN+SAN} 
& {\bf 0.853} & {\bf 0.846} & {\bf 0.852} & {\bf 0.850} & {\bf 0.767} & {\bf 0.761} & {\bf 0.766} & {\bf 0.766}
\\ 
\textbf{Improve} (\%) & 
{\bf +2.8} & {\bf +2.8} & {\bf +1.8} & {\bf +4.1} & {\bf +20.7} & {\bf +30.1} & {\bf +52.3} & {\bf +7.8} \\ \hline    
UPSR 
& 
0.635 & 0.546 & 0.499 & 0.612 & 0.587 & 0.556 & 0.505 & 0.621  \\
\textbf{UPSR+SAN}   
& 
{\bf 0.767} & {\bf 0.758} & {\bf 0.759} & {\bf 0.773} & {\bf 0.757} & {\bf 0.751} & {\bf 0.758} & {\bf 0.756} \\ 
\textbf{Improve} (\%) &  
{\bf +13.2} & {\bf +21.3} & {\bf +26.0} & {\bf +16.1} & {\bf +17.0} & {\bf +19.5} & {\bf +25.3} & {\bf +13.4}
\\ \hline
\end{tabular}
}
\caption{Results of fake news detection for general cold-start propagation. ma-F1 means macro-average F1 score.
The improvements over the corresponding baseline are significant at level p$<$0.05 based on $t$-test. We highlight them in bold.
}
\label{tab:results_general}
\end{table}   

\begin{table}[t]
\centering
\resizebox{0.94\linewidth}{!}{
\begin{tabular}{l|c|c|c|c|c|c|c}
\hline 
\multicolumn{1}{c|}{\multirow{2}{*}{Model}}   & \multirow{2}{*}{Acc.} & \multicolumn{5}{c|}{weighted-F1} &  \multirow{2}{*}{Avg.}  \\ \cline{3-7}
& & \textit{ch} & \textit{fg} & \textit{gc} &\textit{os} & \textit{ss} & \\
\hline 
\multicolumn{1}{l|}{Content} &     
0.538 & 0.367 & 0.507 & 0.582 & 0.671 & 0.523 & 0.530
\\  \hline
GAT 
& 0.644 & 0.503 & 0.492 & 0.733 & 0.643 & 0.466 & 0.567 \\ 
\textbf{GAT+SAN} 
&
{\bf 0.666} & 0.503 & {\bf 0.525} & 0.733 & {\bf 0.672} & {\bf 0.508} & {\bf 0.609} \\
\textbf{Improve} (\%) 
& 
{\bf +2.2} & 0.0 & {\bf +3.3} & 0.0 & {\bf +2.9} &{\bf +4.2} & {\bf +4.2} \\ \hline
GCNFN 
& 
0.633 & 0.503 & 0.580 & 0.728 & 0.594 & 0.546 & 0.590  \\ 
\textbf{GCNFN+SAN} 
&  {\bf 0.659} & {0.503} & {\bf 0.587} & {\bf 0.746} & {\bf 0.678} & {0.472} & {\bf 0.597}  \\ 
\textbf{Improve} (\%) 
&{\bf +2.6} & 0.0 & {\bf +0.7} & {\bf +1.7} & {\bf +8.4} & {-7.4} & {\bf +0.7}  \\     \hline
BiGCN & 
0.620 & 0.468 & 0.471 & 0.727 & 0.658 & 0.324 & 0.529
\\ 
\textbf{BiGCN+SAN} &
{\bf 0.641} & {\bf 0.503} & {\bf 0.563} & {\bf 0.733} & {\bf 0.681} & {\bf 0.428} & {\bf 0.581}
\\ 
\textbf{Improve} (\%) & 
{\bf +2.2} & {\bf +3.6} & {\bf +9.2} &{\bf +0.6} & {\bf +2.3} & {\bf +10.4} & {\bf +5.2}
\\ \hline    
UPSR  &
0.630 & 0.444 & 0.521 & 0.742 & 0.671 & 0.534 & 0.582 \\
\textbf{UPSR+SAN} & 
{\bf 0.664} & {\bf 0.503} & {\bf 0.564} & {0.721} & {0.665} & {\bf 0.578} & {\bf 0.606}
\\ 
\textbf{Improve} (\%) & 
{\bf +3.4} & {\bf +5.9} & {\bf +4.3} & -2.0 & {-0.6} & {\bf +4.4} & {\bf +2.4} \\ \hline
\end{tabular}
}
\caption{Results of fake news detection for  event-level cold-start propagation.  Avg. refers to an average of weighted-F1 of five events.
The bolded improvements over the corresponding baseline model are significant at level p$<$0.05 based on $t$-test.
}
\label{tab:results_general:event}
\end{table}

\subsection{Effects of SAN Framework}
Table~\ref{tab:results_general} and Table~\ref{tab:results_general:event} summarize results of SAN applied to propagation-based methods for general and event-aware cold-start fake news detection tasks. 
Methods that apply the SAN framework consistently outperform the corresponding baseline on all datasets for general and event-aware cold-start fake news detection, which shows the effectiveness of SAN.  

\section{Conclusion}
This paper focuses on the generalization of absence of the whole propagation and explores a new practical fake news detection task towards cold-start propagation, aiming to identify the content-only news by exploiting previously contents and propagation. 
For the task, we design a simple but effective SAN framework to transfer the propagation patterns to the content-only samples. 
Experiments show the poor generalization of existing propagation-based models and the effectiveness of SAN for two types of cold-start propagation.

\section*{Acknowledgements} 
The authors thank anonymous reviewers for their helpful comments.
This work is supported by the National Natural Science Foundation of China (No.6210071416),  the National Key Research and Development Program of China (No. 2022YFC3302102), 
and the National Funded Postdoctoral Researcher Program of China (No. GZC20232969).

\bibliographystyle{IEEEbib}
\bibliography{refs}
 
\end{document}